\documentclass[useAMS,usenatbib,usegraphicx]{mn2e}
\usepackage{times}
\usepackage{amsmath}
\usepackage{amssymb}

\input{psfig.sty}

\title[Star formation in extreme CRDRs]
{Extreme Cosmic-Ray-Dominated-Regions: a new paradigm for high star
 formation density events in the Universe}
\author[Papadopoulos et al.]
  {Padeli P. Papadopoulos,$^{1}$
  Wing-Fai~Thi,$^{2,3}$ Francesco~Miniati,$^4$ Serena~Viti,$^5$\\
  $^1$Argelander-Institut f\"ur Astronomie,
      Auf dem H\"ugel 71, D-53121 Bonn, Germany\\
  $^2$UJF-Grenoble 1 / CNRS-INSU, Institut de Plan\'etologie et d'Astrophysique de
      Grenoble (IPAG) UMR 5274, Grenoble, F-38041, France\\
  $^3$SUPA, Institute for Astronomy, Royal Observatory of Edinburgh,
      University of Edinburgh, Blackford Hill, Edinburgh EH9 3HJ, UK\\
  $^4$Institut f\"ur Astronomie, ETH Z\"urich, 8093 Zurich, Switzerland\\
  $^5$Department of Physics and Astronomy, University College London, London WC1E 6BT,  UK}

\date{Accepted ... ; Received ... ; in original form ...}

\pagerange{000--000}

\begin{document}

\label{firstpage}

\maketitle

\begin{abstract}
We   examine   in   detail    the   recent   proposal   that   extreme
Cosmic-Ray-Dominated-Regions (CRDRs) characterize  the ISM of galaxies
during events  of high-density star  formation, fundamentally altering
its  initial  conditions  (Papadopoulos  2010).  Solving  the  coupled
chemical and thermal state equations for dense UV-shielded gas reveals
that  the large  cosmic ray  energy  densities in  such systems  ($\rm
U_{CR}$$\sim  $  few$\times$(10$^3$--10$^4$)\,$\rm  U_{CR,Gal}$)  will
indeed raise the minimum temperature  of this phase (where the initial
conditions of  star formation  are set) from  $\sim $10\,K (as  in the
Milky Way) to  $\sim $50--100\,K.  Moreover in such  extreme CRDRs the
gas temperature  remains fully decoupled  from that of the  dust, with
$\rm  T_{kin}$$\gg$$\rm  T_{dust}$,   even  at  high  densities  ($\rm
n(H_2)$$\sim$10$^5$--10$^6$\,cm$^{-3}$),  quite  unlike  CRDRs in  the
Milky Way  where $\rm T_k$$\sim $$\rm T_{dust}$  when $\rm n(H_2)$$\ga
$10$^5$\,cm$^{-3}$.   These   dramatically  different  star  formation
initial conditions  will: a) boost  the Jeans mass of  UV-shielded gas
regions by factors of $\sim$10--100 with respect to those in quiescent
or less extreme  star forming systems, and b)  ``erase'' the so-called
inflection point of the effective equation of state (EOS) of molecular
gas.   Both these  effects occur  across the  entire density  range of
typical molecular  clouds, and may  represent {\it a new  paradigm for
all high-density star  formation in the Universe, with  cosmic rays as
the  key driving mechanism},  operating efficiently  even in  the high
dust  extinction  environments  of  compact extreme  starbursts.   The
characteristic  mass of  young  stars  will be  boosted  as a  result,
naturally yielding a top-heavy stellar initial mass function (IMF) and
a bimodal  star formation  mode (with the  occurance of  extreme CRDRs
setting the  branching point).   Such CRDRs will  be present  in local
Ultra Luminous  Infrared Galaxies (ULIRGs)  and merger-driven gas-rich
starbursts across  the Universe where  large amounts of  molecular gas
rapidly dissipate towards compact  disk configurations where they fuel
intense   starbursts.   In   hierarchical   galaxy  formation   models
CR-controlled  SF initial  conditions lend  a physical  basis  for the
currently   postulated   bimodal   IMF  in   merger/starburst   versus
quiescent/disk star  forming environments, while  naturally making the
integrated  galactic IMFs  (IGIMFs) a  function of  the SF  history of
galaxies.

\end{abstract}

\begin{keywords}
 cosmic rays\ -- dust, extinction\ -- galaxies: starburst\ -- galaxies: star formation
\ -- galaxies: interaction\ -- ISM: molecules\ -- ISM: supernovae remants.
\end{keywords}

\section{Introduction}

  Much  of  the stellar  mass  in  the  Universe forms  in  starbursts
  (e.g. Blain et al.  1999; Genzel  et al.  2001; Smail et al.  2002),
  spectacular events during  which the star formation rate  (SFR) of a
  galaxy rises from few solar masses per year, typical of spirals such
  as the Milky-Way (Mckee \&  Williams 1997), to several hundred solar
  masses per  year as  in the local  Ultra Luminous  Infrared Galaxies
  (ULIRGs) (e.g. Sanders \& Mirabel  1996; Genzel et al. 1998; Sanders
  \& Ishida  2004) where merger-driven  starbursts take place  in very
  compact  (D$\sim $100--300\,pc)  dense  gas disks  (e.g.  Downes  \&
  Solomon  1998; Sakamoto  et al.   2008).   In the  so called  sub-mm
  galaxies (SMGs),  similarly merger-driven starbursts  in the distant
  Universe, SFRs  can even  reach thousands of  solar masses  per year
  (Smail  et al.   1997; Hughes  et al.   1998; Eales  et  al.  1999).
  While  stars form  invariably out  of molecular  gas  throughout the
  Universe (e.g.  Solomon et al. 1997; Frayer et al. 1998, 1999; Greve
  et  al.   2005,  Tacconi  et  al.   2006), it  is  in  its  densest,
  UV-shielded, and Cosmic-Ray-Dominated phase where the star formation
  initial  conditions, and  the  characteristic mass  of the  emergent
  young stars  $\rm M^{(*)}  _{ch}$ are truly  set (Bergin  \& Tafalla
  2007;     Elmegreen     et     al.     2008).      Indeed,     while
  Photon-Dominated-Regions  (PDRs) contain the  bulk of  the molecular
  gas mass in  galaxies (Hollenbach \& Tielens 1999;  Ossenkopf et al.
  2007), their  widely varying  physical conditions remain  always far
  removed from those prevailing  in UV-shielded CR-dominated dense gas
  cores.  The location  of the latter deep inside  molecular clouds, a
  strong  temperature regulation  via  the onset  of gas-dust  thermal
  equillibrium at densities  $\rm n(H_2)$$>$10$^4$\,cm$^{-3}$, and the
  nearly complete  dissipation of supersonic  turbulence (Larson 2005;
  Bergin \& Tafalla 2007; Jappsen et  al.  2005), all help to keep the
  thermal state of typical CRDRs a near invariant over a wide range of
  ISM environments.

Cosmic rays  (CRs), accelerated  in shocks around  supernovae remnants
(SNRs) or winds associated with O,  B star clusters (e.g. Binns et al.
2008), are responsible for the heating, ionization level, and chemical
state  of such  dense UV-shielded  regions in  molecular  clouds (e.g.
Goldmith  \&  Langer 1978;  Goldsmith  2001).   Moreover, while  their
heating  contribution  relative  to  that  from  the  (far-UV)-induced
photoelectric effect in PDRs, or  turbulence, can vary within a galaxy
and individual  molecular clouds,  CRs remain responsible  for setting
the  {\it   minimum}  temperature   attainable  in  the   gaseous  ISM
(e.g. Goldmith  \& Langer 1978;  Bergin \& Tafalla 2007;  Elmegreen et
al. 2008).   For the Galaxy  $\rm T_{k}$(min)$\sim $10\,K,  reached in
UV-shielded cores with $\rm n(H_2)$$\sim $($10^4$--$10^6$)\,cm$^{-3}$.

A recent  study has shown that  the average CR  energy densities ($\rm
U_{CR}$) in compact starbursts  typical in ULIRGs (recently found also
in an SMG  at z$\sim $2.3, Swinbank et al.  2010),  will be boosted by
the   tremendeous   factors   of   $\rm   U_{CR}$$\sim   $(few)$\times
$(10$^3$--10$^4$)\,$\rm  U_{CR,Gal}$,  potentially transforming  their
ISM  into  extreme Cosmic-Ray-Dominated  Regions  (CRDRs).  This  will
dramatically  alter the  thermal and  ionization state  of UV-shielded
cores, and  thus the initial conditions  of star-formation, throughout
their     considerable     molecular     gas     reservoirs     ($\sim
$10$^9$-10$^{10}$\,M$_{\odot}$).   Higher   characteristic  mass  $\rm
M^{(*)} _{ch}$  for the young  stars formed in such  ISM environments,
and longer ambipolar diffusion timescales for the CR-heated UV-shieled
cores could thus be a  new paradigm for high-density star formation in
the  Universe  (see  Papadopoulos  2010  for details).   It  is  worth
mentioning  that  recently  such  large  $\rm  U_{CR}$  boosts  ($\sim
10^{3.5}$$\times $U$_{\rm  CR,Gal}$) have  been found in  the Orion\,A
bar, and while  not representing the average SF  initial conditions in
the Galaxy,  they do allow  local studies of  the effects of  large CR
energy densities on the ISM (Pellegrini et al 2009).  Finally CRs have
been shown to  effect the thermal state of  primordial gas, catalyzing
H$_2$ formation  in the absence of  dust ( Jasche,  Ciardi, \& Ensslin
2007).

In this work we solve the coupled chemical and thermal state equations
for UV-shielded gas and its  concomitant dust in extreme CRDRs without
the approximations  employed by Papadopoulos (2010).  We  then use the
thermal states computed for such a gas phase to examine the effects on
$\rm  M^{(*)}  _{ch}$  and  the  so-called  inflection  point  of  the
Effective Equation  of State (EOS),  set at gas densities  above which
$\rm T_{kin}$$\sim  $$\rm T_{dust}$ (e.g. Spaans \&  Silk 2000; Larson
2005 and  references therein).  Both of  these play a  crucial role in
determining the stellar initial mass function (IMF) in galaxies, which
could  thus turn  out to  be very  different in  the CRDRs  of extreme
starbursts. Throughout  this work we  make the standard  assumption of
$\rm U_{CR}$$\propto  $$\rm \dot\rho_{sfr}$ (SFR {\it  density}) as in
Papadopoulos  2010, though  we now  discuss possible  caveats (section
4.4).   Finally, the  different IMFs  expected in  extreme  CRDRs will
modify  the total  SFR and  $\rm \dot\rho  _{sfr}$ values  computed in
galaxies (from e.g.  cm non-thermal  or IR thermal continuum). In this
new context  and throughout  this work SFR  and $\rm  \dot\rho _{sfr}$
will now  denote these  quantities only for  the massive  stars (which
define $\rm U_{CR}$  and power the IR/cm continuum of the ISM), with
their total  values necessarily remaining uncertain,  depending on the
exact IMF, which will no longer be an invariant.

\section{The minimum gas temperature in CRDR$_{\rm s}$}

In this section  we compute the equillibrium gas  temperatures for the
UV-shielded regions  of molecular clouds immersed in  the large cosmic
ray energy densities of the  extreme CRDRs expected in starbursts with
high SFR densities.

\subsection{A simple method}

%
%
\begin{figure}
\centerline{\psfig{figure=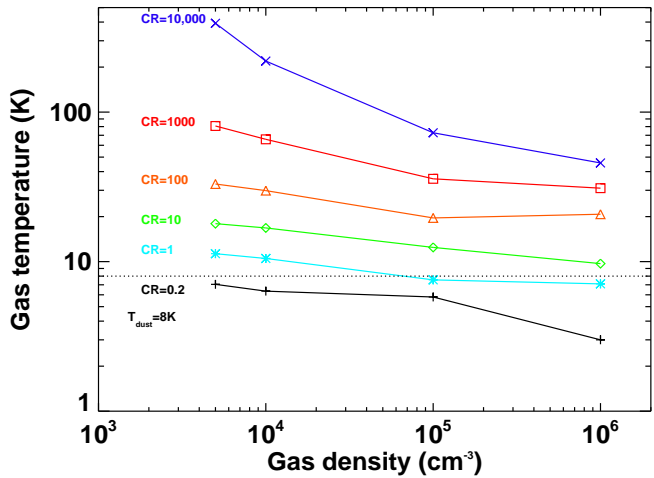,width=3.8in,angle=90}}
\caption{The temperature of UV-shielded  dense gas cores computed from
 Equation 2 for the range of densities typical in
molecular clouds. The CR energy densities range  from the
quiescent environment of the Galaxy ($\rm U_{CR}$=(0.2--1)$\times
$$\rm  U_{CR,Gal}$) to those  in  extreme CRDRs 
 ($\rm   U_{CR}$$\sim $(10$^3$--10$^4$)$\times  $$\rm U_{CR,Gal}$, Papadopoulos 2010). 
  A  dust temperature $\rm T_{dust}$=8\,K  and a Galactic
 CR ionization rate of $\rm \zeta _{CR}$=5$\times $10$^{-17}$\,s$^{-1}$
 have been assumed.}
\label{fig:pdr}
\end{figure}

We  first consider the  simple case  where the  chemical state  of the
UV-shielded dense gas  cores remains unaltered and the  only effect of
large $\rm U_{CR}$ values is to  heat them much more than in quiescent
ISM environments.   Following standard practice  (e.g.  Goldmith 2001)
the  temperature of  such regions  can be  estimated from  the thermal
balance equation

\begin{equation}
\rm \Gamma _{CR}  = \Lambda _{line}  + \Lambda _{gd},
\end{equation}

\noindent
 where $\rm  \Gamma _{CR}\propto  \zeta_{CR}\,n(H_2)$ is the  CR heating
with  $\rm \zeta  _{CR}$$\propto $$\rm  U_{CR}$ being  the  cosmic ray
ionization rate.  The cooling  term $\rm \Lambda _{line}$ is dominated
by CO and  other molecular lines, while $\rm  \Lambda _{gd}$ accounts
for gas cooling due to the gas-dust interaction.

 A  significant  point of  departure  from  the  previous analysis  by
Papadopoulos  (2010) is  the inclusion  of the  atomic  fine structure
lines OI  (at 63\,$\mu $m) and  CII (at 158\,$\mu $m)  in $\rm \Lambda
_{line}$.   Indeed for  the  CR-boosted temperatures  expected in  the
UV-shielded dense  gas of extreme CRDRs, and  unlike the corresponding
gas phase  in the quiescent  ISM of the  Galaxy, the cooling  from the
fine structure atomic line of OI (63$\mu $m) can also become important
and  along with  the gas-dust  interaction overtake  the CO  lines for
typical  core densities of  $\rm n(H_2)$$\geq  $10$^5$\,cm$^{-3}$ when
$\rm T_k$$>$50\,K .  Finally ISM-CR interactions can generate internal
UV radiation which destroys CO,  producing traces of CII, thus cooling
via its  fine structure  line at 158\,$\mu  $m must also  be included.
The thermal balance equation  encompassing all physical processes that
can be important in the UV-shielded  ISM of extreme CRDRs thus~is:

\begin{eqnarray}
\rm \Lambda _{CO}(T_{k}) + \Lambda _{gd}(T_{k},T_{dust})+ 
 \Lambda _{OI\,63}(T_k) + \Lambda _{CII}(T_k)=\Gamma _{CR},
\end{eqnarray} 

\noindent
(for the expressions  used see the Appendix). The  solutions are shown
 in  Figure~1   from  where  it  becomes  obvious   that  the  minimum
 temperatures  attainable  in dense  gas  cores  in  the ISM  of  such
 starbursts  are much  higher  than  in the  Galaxy  or galaxies  with
 moderate $\rm \dot\rho_{sfr}$ values.   Turbulent gas heating was not
 considered since  we want to  compute the minimum  temperature values
 for  UV-shielded  gas  and  since  supersonic  turbulence  (and  thus
 shock-heating) has usually dissipated  in dense pre-stellar gas cores
 where the initial conditions of SF are set.

\begin{figure*}
\centerline{\psfig{figure=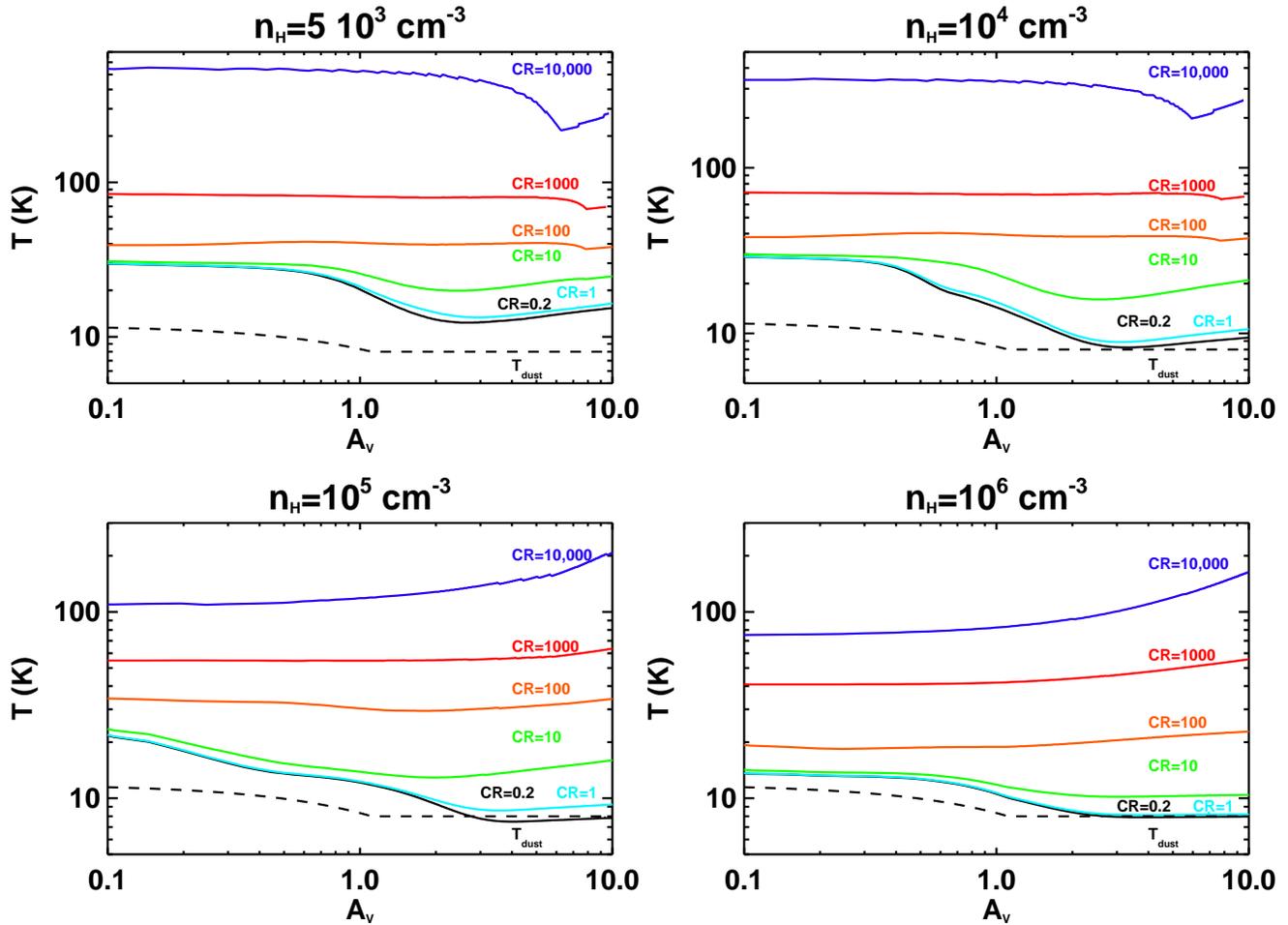,height=5.1in,angle=90}}
\caption{The gas temperature profiles versus optical extinction inside
molecular clouds immersed in a range of CR energy densities, estimated
using  the time-dependant UCL-PDR  code solving  both for  thermal and
chemical  equillibrium at  each point  of  the cloud.   The CR  energy
densities  range  from Galactic  values  $\sim $(0.2--1)$\times  $$\rm
U_{CR,Gal}$, to low-level star forming systems $\sim $10$\times
$$\rm      U_{CR,Gal}$,     and      extreme      starbursts     $\sim
$(10$^3$--10$^4$)$\times  $$\rm U_{CR,Gal}$.  The  densities encompass
the range typical for  star-forming molecular clouds, and the incident
far-UV field assumed is $\rm G_{\circ}$=1 (in Habing units).  The dust
temperature  $\rm T_{dust}(A_v)$  profile is  indicated by  the dashed
black line.  Stronger radiation  fields reaching the dense pre-stellar
cores  deep  inside molecular  clouds  (e.g.   because of  ``leaking''
far-UV radiation from O,B  stars inside the typically clumpy molecular
ISM), can  further raise the  gas temperatures for $\rm  A_v$$\leq $5,
while    deeper   than   that    gas   temperatures    remain   always
CR-controlled. The small  $\sim $10\% rise of $\rm  T_{k}(A_v)$ at the
highest  densities and  $\rm U_{CR}$  values for  $\rm A_v$$\ga  $5 is
likely an  artifact of  the high optical  depths of  C and CO  at such
large cloud depths (see Bayet et al. 2010).}
\label{fig:T_vs_Av}
\end{figure*}

From  Figure  1 it  can  be seen  that  the  minimum gas  temperatures
 possible  in   extreme  CRDRs  are  lower  than   those  computed  by
 Papadopoulos (2010),  a result of  the extra cooling  power emanating
 from the now  included atomic transitions that become  excited at the
 high temperatures now possible in an ISM permeated by large CR energy
 densities.   Nevertheless  these  minimum possible  gas  temperatures
 remain $\sim $5--10  times higher in extreme CRDRs  than those in the
 Milky  Way or galaxies  with moderate  $\rm \dot\rho  _{sfr}$ values.
 Any  IR radiation  ``leaking''  deep  in the  cloud  and warming  the
 concomitant dust  above $\sim  $(8--10)\,K, or any  remnant turbulent
 gas  heating  (Pan \&  Padoan  2010)  can  only raise  these  minimum
 temperatures further.

\subsection{Solving for the coupled chemical and thermal state
 of UV-shielded gas cores in extreme CRDR$_{\rm S}$}

The thermal balance equation  for UV-shielded, CR-heated gas (Equation
2) and its solutions (Figure~1), while physically transparent, they do
not account for the fact that ISM chemistry (and thus the abundance of
coolants) is  also strongly CR-regulated in such  regions.  This makes
the  problems of  thermal  and chemical  balance  of CR-dominated  gas
strongly coupled  that must be  solved together, especially  given the
very  large CR  energy  densities  possible in  the  CRDRs of  compact
starbursts.  We do so by utilizing the advanced time-dependant UCL-PDR
code (the  latest version described in  Bell et al.   2007) to compute
gas    temperatures   of    CR-heated   molecular    cloud   interiors
self-consistently with  their CR-regulated chemistry for  a wide range
of CR energy densities. This  model has already been successfully used
several times to model PDRs in the Galaxy (e.g. Thi et al. 2009).

The  heating mechanisms  included are:  (i) photoelectric  effect from
silicate  grains  and Polycyclic  Aromatic  Hydrocarbons (PAHs),  (ii)
H$_{2}$ formation  on gain surfaces,  (iii) H$_{2}$ photodissociation,
(iv) H$_{2}$ UV fluorescence,  (v) CII recombination, (vi) interaction
of  low-energy cosmic  rays with  the  gas.  The  latter dominate  gas
heating deep inside  H$_2$ clouds ($\rm A_v$$\geq $5)  where dense gas
cores reside  and the initial  conditions for star formation  are set.
At cloud  surfaces hot  gas emits Ly$\alpha$  and OI\,63$\mu  $m lines
while  deeper   in  ($\rm   A_{v}$$>$0.1),  gas  cooling   occurs  via
fine-structure line emission of  oxygen and neutral and singly ionized
carbon.  Vibrational  and ro-vibrational lines of H$_{2}$  and CO also
cool the gas while CO  rotational lines are excellent coolants for the
cold  inner regions  as the  first rotationally-excited  level reaches
down to  5.5~K.  The interaction between  the gas and  the dust grains
acts  as  cooling  or heating  agent  for  the  gas depending  on  the
difference  between the  gas  and dust  temperature.   H$_{2}$ and  CO
self-shielding are  taken into account.   The code calculates  the gas
chemical  abundances, emergent cooling  line fluxes,  and temperatures
self-consistently  at  each  depth  in  the cloud,  and  it  has  been
benchmarked  against other  state-of-the art  photodissociation region
codes in R\"ollig, M.  et al.  (2007).

We set a nominal incident  far-UV field of $\rm G_{\circ}$=1 in Habing
units (expected to be even  lower for the dense pre-stellar cores deep
inside H$_2$ clouds) and a residual micro-turbulent line width of $\rm
\Delta  V_{m-turb}$=5\,km\,s$^{-1}$ (which  enters the  expressions of
line optical depths).  In  all cases chemical and thermal equillibrium
is  reached fast,  within  T$_{\rm chem,th}$$\leq  10^4$\,yrs, and  is
always  shorter than the  local dynamical  timescale which  drives the
ambient star formation events and  thus the temporal evolution of $\rm
U_{CR}$.  The results,  shown in  Figure~2, corroborate  the  high gas
temperatures  obtained using  the  simpler method.   This reveals  the
cosmic  rays  as {\it  the  most  important  feedback aspect  of  star
formation}  capable of significantly  altering its  initial conditions
during events of high star formation density via the strong heating of
the  deeply embedded and  UV-shielded dense  gas regions  of molecular
clouds.  It  can thus operate  unhindered in the high  dust extinction
environments  of  extreme starbursts  such  as  local  ULIRGs or  high
redshift SMGs.

\subsubsection{Testing the results for stronger radiation fields}

We also ran the same grid of models with the incident far-UV radiation
field scaling exactly as the  CR energy densities which corresponds to
the maximum  incident radiation field  that can be expected  for dense
gas cores  (in practice pre-stellar cores are  always UV-shielded deep
inside molecular clouds).  The  dust temperature is now more important
than in  the previous  case of $\rm  G_{\circ}$=1, and is  computed as
part  of the  code.  Its value  inside  the cloud  can influence:  the
cooling  by emitting  IR photons  which interact  with  line radiative
transfer, the  H$_2$ formation  by changing the  sticking probability,
the  evaporation/condensation of molecules,  and most  importantly the
gas-grain heating/cooling. The UCL-PDR code uses a modified version of
the  formula of  Hollenbach, Takahashi,  \& Tielens  (1991) (hereafter
HTT) corrected for the various mean grain sizes.

The HTT  formula has been  further modified to include  attenuation of
the  far-IR radiation.   The incident  FUV radiation  is  absorbed and
re-emitted in the infrared by dust  at the surface of the cloud (up to
$\rm  A_v$$\sim$1\,mag).  In  the HTT  derivation, this  FIR radiation
then  serves  as  a  second  heat  source for  dust  deeper  into  the
cloud. However  in their model  this secondary radiation  component is
not attenuated  with distance  into the cloud,  remaining undiminished
with  depth and  leading to  higher  dust temperatures  deep into  the
cloud,    which   in    turn    heats   the    gas   to    unrealistic
temperatures. Attenuation of the  FIR radiation has been introduced by
using  an approximation  for the  infrared-only dust  temperature from
Rowan-Robinson    (1980),    his    Equation   30b:    $\rm    T$=$\rm
T_{\circ}(r/r_{\circ})^{-0.4}$  where  $\rm  r_{\circ}$ is  the  cloud
depth corresponding  to $\rm  A_v$=1\,mag outer 'slab'  that processes
the incident FUV radiation and then re-emits it in the FIR.

From Figure 3 it can be  readily seen that the gas temperatures remain
invariant  beyond $\rm  A_v$$\sim $5  even with  with  incident far-UV
fields  as enhanced  as  the CR  energy  densities.  This  essentially
leaves CRs {\it  as the only strong feedback  factor of star formation
operating  efficiently  deep   inside  molecular  clouds.}   Moreover,
intense CR heating of UV-shielded  gas strongly decouples gas and dust
temperature even  at high densities  (Figures 1, 2, 3)  with important
implications  on the  gas  fragmentation properties  and  thus on  the
emergent stellar IMF (section 3.2).

\begin{figure*}
\centerline{\psfig{figure=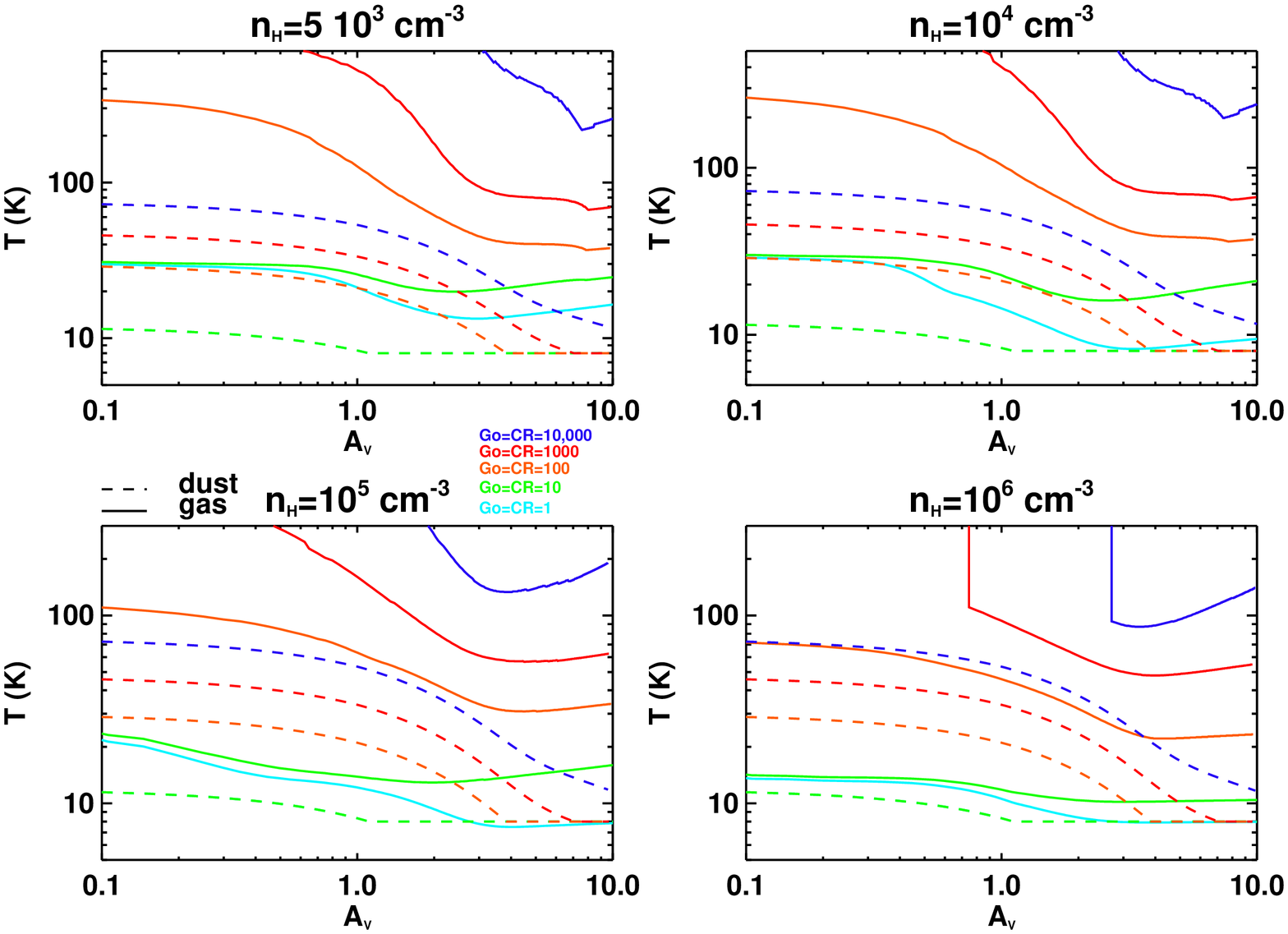,height=5.1in,angle=90}}
\caption{The computed gas and dust temperature profiles versus optical
extinction inside  molecular clouds as in  Figure 2, but  now with the
incident far-UV radiation field also scaling as the CR energy density.
The now  strongly varying dust  temperatures are shown along  with the
gas  temperatures  as dashed  lines  with  the  same color.   The  gas
temperatures at $\rm A_v$=5--10  remain almost invariant (compare with
Figure 2)  demonstrating the near-invariance  of the thermal  state of
the  gas (and  thus of  the  star formation  initial conditions)  deep
inside  molecular  clouds when  it  comes  to  its dependence  on  the
radiation  field. The  abrupt  $\rm T_{k}$  drop  at high  densities
($\sim$($10^5-$$10^6$)\,cm$^{-3}$) and G$_{\circ}$  values is due to a
lack of  atomic/ion coolants in the  H-rich region: as  H is converted
into H$_2$, the  H$_2$ formation heating (which was  the major heating
agent) drops while at the same time CI cooling (CII is converted to CI
due   to  H$_2$/C   mutual  shielding)   and  H$_2$   cooling  sharply
increases. For high  $\rm U_{CR}$ the gas remain  much warmer than the
dust, marking  a {\it CR-induced  thermal decoupling of gas  and dust}
(section~3.2).}
\label{fig:T_vs_Av_CR=Go}
\end{figure*}

\section{New initial conditions for star formation, and the stellar IMF in extreme CRDR$_{\rm S}$}

 The new detailed  thermo-chemical calculations presented here support
 the results from the previous study  of CRDRs and the effects of high
 $\rm  U_{CR}$ values  on  the initial  conditions  of star  formation
 (Papadopoulos 2010).   The thermal state of UV-shielded  dense gas is
 indeed dramatically  altered in the CRDRs of  extreme starbursts, and
 this will  affect its Jeans  mass, gas fragmentation  properties, and
 the  charecteristic mass $\rm  M^{(*)} _{ch}$  of young  stars.  This
 contradicts  the results  by Elemgreen  et al.   (2008)  which, while
 correctly considering the UV-shielded gas  phase as the one where the
 SF  initial conditions  are  set,  it omitted  CRs  from its  thermal
 balance.

 It must be pointed out  that while several other studies investigated
 the effects  of starburst environments  on the IMF, all  (besides the
 one by Elmegreen et al.  2008) failed to consider the ISM phase where
 the initial conditions  of SF are actually set  (the CRDRs).  Instead
 they used physical  properties typical of PDRs (e.g.   Klessen et al.
 2007) which, while  dominating the  observed dust SEDs  and molecular
 line  Spectral Energy  Distributions  (SLEDs) of  galaxies, they  are
 hardly representative  of the  physical conditions expected  in their
 CRDRs. Even  for mildly SF IR-luminous galaxies  the dust temperature
 typical of  their global  SEDs is $\sim  $(30--40)\,K, which  is well
 above $\rm T_{dust}$$\sim $(8--10)\,K expected in their CRDRs.  Using
 inappropriate  PDR-type  initial  conditions  of  star  formation  in
 numerical models: a) would make them hard-pressed to account even for
 the  near-invariant IMF found  in the  Galaxy (as  the wide  range of
 physical conditions in  its massive PDR gas phase  would yield a wide
 range  of  IMFs),  b)  omits  the  significant,  CR-induced,  thermal
 decoupling  between  gas  and   dust  (Figures  2,  3).   The  latter
 fundamentally  impacts the  so-called inflection  point  of effective
 equation of  state (EOS) used  in all realistic (i.e  non isothermal)
 numerical simulations of  molecular cloud fragmentation (e.g. Jappsen
 et al. 2005;  Bonnell et al. 2006), and this  is further discussed in
 section 3.2.

\subsection{The characteristic mass of young stars in  CRDRs}

The thermal state of these dense  pre-stellar gas cores is no longer a
near-invariant but is  instead strongly altered by the  high CR energy
density  backgrounds of extreme  starbursts, inevitably  affecting the
fragmentation of  those gas  regions towards protostars.   Indeed, the
Jeans mass for such CR-heated cores

\begin{eqnarray}
\rm \nonumber M^{(c)} _{J}&=&\rm \left(\frac{k_B T_k}{G\mu m_{H_2}}\right)^{3/2} \rho_c ^{-1/2}=\\
&=&\rm 0.9\left(\frac{T_k}{10K}\right)^{3/2}\left[\frac{n_c(H_2)}{10^4 
cm^{-3}}\right]^{-1/2}\, M_{\odot},
\end{eqnarray}

\noindent
 rises  from  $\rm  M^{(c)}  _J$$\sim  $0.3\,M$_{\odot  }$  when  $\rm
  n_c$(H$_2$)=10$^5$\,cm$^{-3}$ and  $\rm T_k$=10\,K (typical  for the
  Galaxy),  to $\rm  M^{(c)} _J$$\sim  $(3--10)\,M$_{\odot}$  for $\rm
  T_k$$\sim $(50--110)\,K expected for same density H$_2$ gas immersed
  in the  very large CR  energy densities of extreme  starbursts.  The
  near-invariance of  $\rm M^{(c)} _J$  in galaxies (Elmegreen  et al.
  2008) is  thus upended in  starbursts with  high SFR  densities.  In
  Figure 4 the $\rm M_{J}$  values computed for molecular gas immersed
  in  the CRDRs  of compact  extreme starbursts  are shown  for  a gas
  density range  typical for star-forming molecular  clouds.  For $\rm
  U_{CR}/U_{CR,Gal}$$\geq $10$^3$,  $\rm M_{J}$ increases  by a factor
  of $\sim $10 across the  entire range, which will invariably lead to
  a higher $\rm M^{(*)} _{ch}$ and thus a top-heavy IMF (Larson et al.
  2005; Elmegreen  et al.   2008).  The wide  range of  densities over
  which this occurs makes  this conclusion independent of the specific
  details   of  molecular   cloud  fragmentation   as  long   as  $\rm
  M_J[n(\vec{r},t), T_k(\vec{r}, t)]$ drives the fragmentation process
  outcome at each spatial and  temporal point $\rm (\vec{r}, t)$.  The
  latter is  the current consensus  on the role  of the Jeans  mass in
  molecular  cloud fragmentation  towards  a stellar  IMF, though  the
  views regarding the particular gas phase whose $\rm M_{J}$ sets the
  $\rm M^{(*)} _{ch}$ vary (e.g. Klessen 2004; Bonnell et al.~2006).

\begin{figure}
\centerline{\psfig{figure=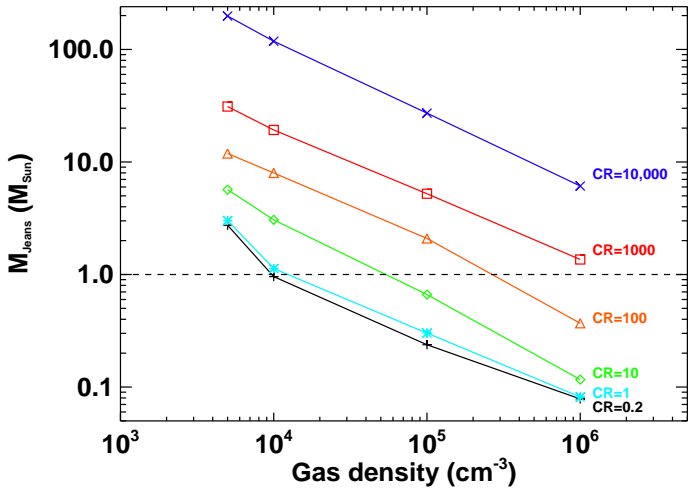,width=3.9in,angle=90}}
\caption{The   Jeans  mass   estimated  from   Equation  3,   and  gas
temperatures  inside CRDRs, computed  for the  corresponding densities
(using  the  full UCL-PDR  code)  and  averaged  over the  inner  $\rm
A_v$=5--10 cloud regions  (Figure 2). Low mass star  formation will be
clearly   surpressed  in   extreme  CRDRs   for  a $\rm M_J$-driven  gas
fragmentation towards a stellar IMF (see text).}
\label{fig:M_jeans}
\end{figure}

  If $\rm  M^{*)} _{ch}$$\propto $$\rm  \langle M_{J}\rangle $  at the
  {\it  onset} of  molecular cloud  fragmentation (Klessen  2004), the
  CR-boost of $\rm M^{(*)} _{ch}$ in extreme CRDRs is obvious from the
  large vertical  displacement of the $\rm M_{J}$  values in Figure~4.
  In  a non-isothermal  gravoturbulent  molecular cloud  fragmentation
  scheme   (e.g.   Jappsen   et  al.    2005)  it   is   $\rm  M^{(*)}
  _{ch}$$\sim$$\rm M_J$($n_c$)  at a characteristic  gas density $n_c$
  above  which  efficient gas-dust  thermal  coupling  lowers the  gas
  temperature   in   CRDRs   towards   a   minimum   value   of   $\rm
  T_{k}$(min)$\sim $$\rm T_{dust}$ (Larson 2005).  For the Galaxy this
  occurs at $ n_c$$\sim  $10$^5$\,cm$^{-3}$, a density that marks also
  the  transition  to  near-isothermal  cloud  regions  where  thermal
  motions dominate over supersonic  turbulence (Larson 2005; Bergin \&
  Tafalla 2007).   From Figures 2,  3 it is  obvious that the  high CR
  energy densities  in extreme CRDRs {\it  will keep the  gas and dust
  thermally  decoupled, with $  T_{k}$(min)$\gg $$\rm  T_{dust}$}, and
  thus a characteristic  density can no longer be  marked by the onset
  of  thermal  equillibrium between  gas  and  dust.  Such  CR-induced
  effects have been long suspected  for the dense molecular gas in the
  Galactic Center (Yusef-Zadeh et al.~2007).

   In  the  CR-innudated  molecular  clouds of  extreme  starbursts  a
   characteristic density  in the ISM can  now remain only  as the one
   where the supersonic turbulence  has fully dissipated and overtaken
   by thermal  motions in pre-stellar cores, a  prerequisite for their
   efficient  gravitational collapse  (e.g. Krumholz  \&  McKee 2005).
   The  average density of  such transition  regions can  be estimated
   from  two well-known  scaling relations  in molecular  clouds: $\rm
   \sigma _V(r)$=$\rm \sigma _{\circ} (r/pc)^{h}$ (linewidth-size) and
   $\rm \langle n \rangle $=$\rm n_{\circ} (r/pc)^{-1}$ (density-size)
   after  setting  $\rm   \sigma  _V$(min)=$\rm  \left(3  k_B  T_k/\mu
   m_{H_2}\right)^{1/2}$  as  the   minimum  linewidth  possible,  and
   solving for the corresponding mean volume~density,

\begin{equation}
\rm \langle n_{tr} \rangle  = n_{\circ} \left(\frac{\mu m_{H_2} \sigma
^{2}       _{\circ}}{3       k_B       T_k}\right)^{1/2h}\sim       11
n_{\circ}\left(\frac{\sigma              _{\circ}}{km\,s^{-1}}\right)^2
\left(\frac{T_k}{10K}\right)^{-1}.
\end{equation}

\noindent
For  h=1/2  (expected  for   gas  clouds  virialiazed  under  external
pressure,  Elemgreen 1989),  $\sigma  _{\circ}$=1.1\,km\,s$^{-1}$, and
for $\rm n_{\circ}$=few$\times  $10$^3$\,cm$^{-3}$, it is $\rm \langle
n_{tr}   \rangle$$\sim   $few$\times   $10$^4$\,cm$^{-3}$.   For   the
CR-boosted gas temperatures in extreme starbursts where typically $\rm
T_k$(min)$\sim $100\,K  (Figures 1, 2, 3):  $\rm \langle n_{tr}\rangle
$$\sim    $few$\times    10^3$\,cm$^{-3}$,    and    the    (turbulent
gas)$\rightarrow$(thermal  core)   transition  will  occur   at  lower
densities {\it  as well as} higher (CR-boosted)  gas temperatures than
in quiescent  CRDRs.  This will  then shift the expected  $\rm M^{(*)}
_{ch}$ both vertically  (higher temperatures), as well as  to the left
(lower  densities) of  Figure  4, yielding  even  higher $\rm  M^{(*)}
_{ch}$ values than those resulting  from only a vertical shift of $\rm
M_{J}$ because of CR-boosted gas temperatures.

Thus  irrespective of  molecular cloud  fragmentation details,  {\it a
much  larger characteristic  mass of  young stars  is expected  in the
CR-innudated  molecular clouds  of extreme  starbursts.}   However the
exact  shape  and mass  scale  of  the new  stellar  IMF  can only  be
determined  with dedicated  molecular cloud  fragmentation simulations
that make use of the new SF initial conditions in extreme CRDRs, along
with an EOS appropriate for  the very different thermal and ionization
state of their molecular gas.

\subsection{CRDRs: SF initial conditions, the EOS inflection point}

The importance of the  Effective Equation of State (EOS), parametrized
as a  polytrope P=$K\rho  ^{\gamma}$, in the  fragmentation properties
and emergent mass spectrum  of turbulent self-gravitating gas has been
well-documented in a  few seminal papers (Li et  al.  2003; Jappsen et
al.   2005; Bonnell et  al.  2006;  Klessen et  al. 2007)  revealing a
crucial role for $\gamma $ in  defining the shape of the collapsed gas
core  mass distribution  (and hence  the stellar  IMF).  The  EOS, its
polytropic  index  $\gamma $  as  a  function  of density,  and  their
dependance  on  ISM  properties  such  as  metallicity,  and  physical
conditions such as  the elevated CMB at high  redshifts have also been
well-documented (Spaans  \& Silk 2000,  2005), with the  so-called EOS
inflection point emerging as a  general characteristic. In an EOS that
seems  well-approximated by  a piecewise  polytrope with  two distinct
$\gamma $ values, the latter is simply the characteristic density $\rm
n_c$ for which $\gamma $$<$1  when n$<$$\rm n_c$, and $\gamma $$\ga $1
for n$\ga  $$\rm n_c$.  Its crucial  role in determining  the IMF mass
scale,  discussed  briefly  in   3.1,  is  documented  throughout  the
literature (Larson  2005; Jappsen et  al.  2005; Elmegreen  2008).  In
starburst environments  a ``sharpening''  of that inflection  point is
thought to occur because of warm  dust.  There $\gamma $ reaches up to
$\sim $1.4  and stays  above unity for  densities past  the inflection
point  $\rm  n_c$$\sim  $(10$^4$--10$^5$)\,cm$^{-3}$ (Spaans  \&  Silk
2005),   which   surpresses   molecular  cloud   fragmentation   while
``tilting'' it towards top heavy  IMFs (e.g.  Li et al.  2003; Klessen
et al.~2007).

 Unfortunately  PDR-like ISM  has  been  used to  set  the SF  initial
conditions for  starbursts in all  of the aforementioned  models, even
though  the dust  in CRDRs  (where  these conditions  are really  set)
remains cool,  much lower than  the gas, especially in  extreme CRDRs.
This insensitivity of the CRDR dust temperatures to the ambient far-UV
radiation field can be demonstrated analytically by simply considering
the location of CRDRs  deep inside far-UV illuminated molecular clouds
at    $\rm    A_v\ga   5$    (e.g.     McKee    1989).    Then    $\rm
T_{dust,SB}/T_{dust,Gal}$$\sim$$\rm
\left[exp\left(-A_v/1.086\right)\times     G_{\circ,     SB}/G_{\circ,
Gal}\right]^{1/(4+a)}$ will be the radiation induced boost of the CRDR
dust temperature in starbursts  $\rm T_{dust,SB}$ with respect to that
in  the  quiescent  Galactic  environments  $\rm  T_{dust,  Gal}$$\sim
$8\,K. For a dust emissivity law of $\alpha$=2, a large enhancement of
the far-UV radiation  field $\rm G_{\circ, SB}/G_{\circ, Gal}$=10$^4$,
and  $\rm Av$=5  it is  $\rm  T_{dust,SB}/T_{dust,Gal}$=2.15, yielding
$\rm T_{dust}$$\sim  $17\,K, in good agreement with  the detailed dust
temperature profiles past $\rm A_v$$\sim  $5 shown in Figure 3.  It is
actually this robustness of dust temperatures inside CRDRs, along with
a strong gas-dust coupling at high densities (when $\rm U_{CR}$ values
are modest) that  helps retain: the thermal states  of the pre-stellar
gas  cores, the  $\rm  M^{(*)} _{ch}$  values,  and the  IMF, as  near
invariants over a  wide range of ambient PDR  conditions (Elmegreen et
al. 2008)

Thus PDR  conditions, while always  dominating the observed  dust SEDs
and molecular SLEDs of even mildly  SF galaxies (as PDRs are warm {\it
and}  contain   the  bulk  of  the   ISM  mass)  {\it   they  are  not
representative of  the SF initial  conditions in galaxies,}  which are
set in CRDRs. This leaves the pre-stellar gas phase in them, amounting
to a few\% of the mass per Giant Molecular Cloud ($\sim $SF efficiency
per  typical  GMC),  yet to  be  explored  in  terms  of its  EOS  and
fragmentation properties.   A detailed  exploration of those  in CRDRs
will be  the subject of  a future paper,  but Figures 2 and  3 already
show that  CRs, by inducing  a strong decoupling  of the gas  and dust
temperatures with  $\rm T_{k}$$\gg$$\rm T_{dust}$,  they ``erase'' any
putative  EOS   inflection  point  for   gas  densities  up   to  $\rm
n$=10$^6$\,cm$^{-3}$.   This  will  maintain efficient  molecular  gas
cooling, with  cooling times always much shorter  than dynamical times
(Figure  5), and  $\gamma$$<$1 in  CRDRs over  the entire  gas density
range explored here. In  such regions gas gravoturbulent fragmentation
will   then   occur   equally   efficiently  from   n$\sim   $$5\times
10^3$\,cm$^{-3}$ up to $\sim  $10$^6$\,cm$^{-3}$ yielding a pure power
law IMF (Spaans \& Silk 2000; Li et al. 2003).

The latter result,  along with our discussion in  3.1 makes clear that
the stellar IMF  will be very different in  extreme CRDRs, though only
detailed  gravoturbulent  simulations  with  the  appropriate  initial
conditions, and performed with a new suitable EOS can decide its exact
shape  and mass  scale.  It  must  be also  noted that  the very  high
ionization fractions in extreme  CRDRs can now ``anchor'' the magnetic
field  for much  longer  periods,  and at  much  higher densities,  as
ambipolar   diffusion  in   such   regions  is   now  much   prolonged
(Papadopoulos 2010).  Magnetic fields may thus remain important during
much later  stages of  gas fragmentation, even  in a  highly turbulent
medium, and  may thus have to  be included in the  simulations of this
process in extreme CRDRs.

\begin{figure}
\centerline{\psfig{figure=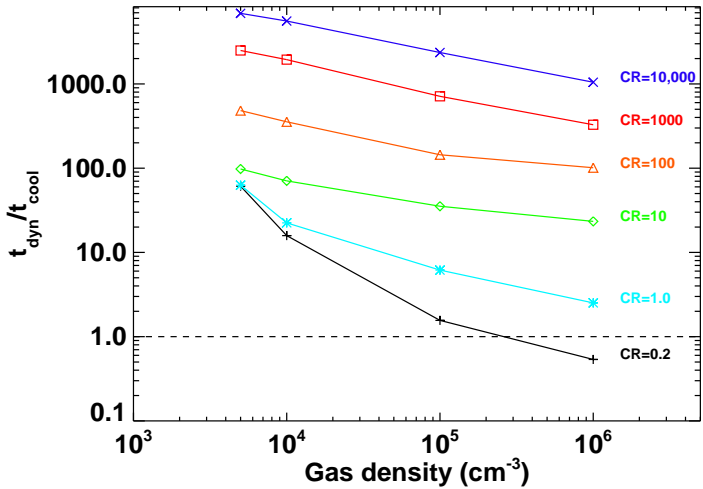,width=3.9in,angle=90}}
\caption{The  ratio  of  dynamical/cooling timescales  in  UV-shielded
molecular gas  of CRDRs ($\rm  A_v$=5-10) estimated using  the UCL-PDR
code and  cosmic ray energy  densities ranging from those  expected in
quiescent Galactic to compact extreme starbursts.}
\label{fig:tdyn_tcool}
\end{figure}

\section{Consequences, observational prospects}

The current  study supports  the recently suggested  role of CRs  as a
decisive  SF  feedback factor,  capable  of  altering  the SF  initial
conditions and the  IMF in extreme CRDRs.  In  this section we discuss
some of  the consequences beyond the  immediate issues of  the ISM and
the stellar IMF  in such enviornments, and outline  current and future
observations that  can shed light  on these new SF  initial conditions
that will be typical in high SFR density events in the Universe.

\subsection{A sequence of $\bf \dot\rho _{sfr}$-dependent stellar IMFs}

The  dependance  of  the   thermo-chemical  and  ionization  state  of
 pre-stellar dense gas in CRDRs on the average $\rm U_{CR}$ permeating
 the ISM of a galaxy naturally links the SF initial conditions and the
 resulting  IMF  to  the  average SFR  density  $\rm  \dot\rho_{sfr}$,
 provided  $\rm  U_{CR}\propto  \dot\rho_{sfr}$.   The details  of  CR
 transport/escape mechanisms in quiescent SF disks and starbursts will
 define  the proportionality  factor  (see Suckhov  et  al.  1993  for
 detailes). Thus,  once simulations of molecular  gas fragmentation in
 CRDRs  produce libraries  of the  corresponding IMFs,  these  will be
 naturally  parametrized  by  $\rm  \dot\rho_{sfr}$, and  hence  actual
 observables  (especially  in  the  upcoming era  of  high-resolution
 imaging at submm wavelengths with ALMA).

\subsection{Implications for hierarchical galaxy formation models: 
a physically-motivated IMF bimodality}


  In hierarchical galaxy formation models a bimodal IMF, top-heavy for
 mergers/starbursts and regular  for disk/regular-SF systems, has been
 postulated     to     explain      the     populations     of     the
 starbursting/merger-driven submillimeter  galaxies (SMGs) and regular
 SF  disks like  the Ly-break  galaxies  (LBGs) at  high redshifts  in
 $\Lambda$CDM cosmology  (Baugh et  al.  2005).  The  CR-controlled SF
 initial conditions  in galaxies,  unhindered by dust  extinction, can
 now  set this on  a firm  physical basis.   From Figures  2 and  4 is
 obvious    that   for   typical    SF   disks    ($\rm   U_{CR}$$\sim
 $(0.2--10)$\times  $$\rm U_{CR,Gal}$) no  substantial effects  on the
 gas temperatures and Jeans masses  in CRDRs are expected while in the
 extreme    CDRDs   of    compact   starbursts    ($\rm   U_{CR}$$\sim
 $(10$^3$--10$^4$)$\times
 $$\rm U_{CR,Gal}$) very warm gas  and large Jeans masses will lead to
 an  effective surpression  of  low-mass star  formation  and hence  a
 top-heavy  IMF.   For merger-driven  star  formation  the latter  may
 characterize  entire star-forming episodes  (i.e.  until  the in-situ
 molecular  gas  reservoir  is  nearly exausted)  as  dissipation  can
 continously  ``funnel''  the  large   molecular  gas  masses  of  the
 progenitors  towards  very  compact ($\sim  $(100--500)\,pc)  regions
 where extreme, Eddington-limited SF,  takes place with tremendeous IR
 brightness      of      $\rm     \sigma      _{IR}$$\sim$(few)$\times
 $10$^{13}$\,L$_{\odot}$\,kpc$^{-2}$ (e.g.  Thompson 2009).  Such high
 IR  brightness,  and thus  high  SFR  densities,  have been  recently
 deduced also in a distant SMG (Swinbank et al. 2010).
 
Thus CR-controlled SF initial  conditions naturally yield a bimodal SF
process in  galaxies, with high $\rm  \dot\rho_{sfr}$ values providing
the branching point, as follows:\\
\noindent
  (merger-driven  SF)$\rightarrow $(high $\rm \dot\rho_{sfr}$)$\rightarrow$(high
    $\rm  U_{CR}$)$\rightarrow $(top heavy IMF),\\
\noindent
(regular disk SF)$\rightarrow $(low $\rm \dot\rho_{sfr}$)$\rightarrow$(low
    $\rm  U_{CR}$)$\rightarrow $(Galactic IMF).\\
\noindent
This proposed $\rm \dot\rho_{sfr}$-IMF  link can then parametrize such
 an SF bimodality  in hierarchical  galaxy formation,  once IMF=F($\rm
 \dot\rho_{sfr}$) ``libraries'' are obtained from e.g.  gravoturbulent
 simulations of  molecular cloud fragmentation  for a range  of $\rm
 \dot\rho_{sfr}$ values.

A top-heavy IMF  in high-$\rm \dot\rho _{sfr}$ systems  will have very
serious  implications   on  the  actual  SFRs  deduced   from  the  IR
luminosities  of these typically  very dust-enshrouded  systems (where
all the far-UV light from massive stars is reprocessed into IR). Their
CR-induced top-heavy IMFs will  now correspond to several times higher
energy outputs  per stellar  mass, and the  SFRs in such  systems will
have to  be revised downwards by  identical factors.

\subsection{A $\rm \dot\rho _{sfr}$-dependent IMF: integrated galactic IMFs
dependent on star formation history?}

A  $\rm  \dot\rho _{sfr}$-dependent  IMF  as  the  natural outcome  of
CR-controlled  SF initial  conditions  in galaxies  can also  naturaly
yield  integrated  galactic IMFs  (IGIMFs)  that  depend  on the  star
formation history  (SFH) of galaxies.   This is because  $\rm \dot\rho
_{sfr}$  can  be a  strong  function of  the  evolution  of a  galaxy,
especially during its early and presumably gas-rich state even when no
mergers are  involved.  Such a  SFH-dependance of the IGIMFs  has been
considered  before as  the  cause of  the well-known  mass-metallicity
relation of galaxies (K\"oppen, Weidner, \& Kroupa 2007).

\subsection{CR propagation in dense molecular gas: a dynamic view is needed}

Setting  $\rm U_{CR}$$\propto$$\rm  \dot\rho_{sfr}$  in CRDRs  assumes
that CRs can freely penetrate and warm the entire volume of a GMC down
to  its densest  UV-shielded regions.   Recent  gamma-ray observations
have directly measured CR energy  densities in the nucleus of a nearby
galaxy on par with those  computed for extreme CRDRs and expected from
its SFR density, while  indicating only mildly calorimetric conditions
for  the CR/ISM  p-p interactions  (Acero  et el.   2009). The  latter
supports an  unhintered CR  penetration through the  molecular clouds.
Nevertheless Alfv\'en  waves generated {\it outside} such  clouds by a
net flux  of CRs in their  environments could be  effective in keeping
low  energy CRs  below  a few  hundred  MeV (those  most effective  in
heating  the gas)  outside  their high  density  regions (Skilling  \&
Strong 1976), which in turn  would affect their thermal balance, their
fragmentation  properties,  and the  emergent  stellar  IMF. Thus  the
coupled  problems  of  CR  transport,  magnetic  field  dynamics,  and
gravoturbulent  evolution of non-isothermal  dense molecular  gas with
high  degrees of  ionization like  those expected  in  compact extreme
starbursts  such  as Arp\,220  (e.g.   Greve  et  al.  2009)  must  be
investigated with dedicated MHD simulations.

\subsection{Observational prospects}

The  detections of $\gamma  $-ray emission  from two  nearby galaxies,
namely M\,82  and the  starburst nucleus of  NGC\,253 (Acciari  et al.
2009, Acero et al.  2009),  are two very important recent developments
in $\gamma $-ray astronomy,  given that $\gamma $-ray emission results
from CR proton interactions with  the molecular ISM.  This allows such
observations to directly measure the average CR energy densities in SF
environments, and  thus sensitive such measurements in  ULIRGs such as
the nearby ULIRG Arp\,220 will  be very important in revealing extreme
CRDRs in such systems.

On the other hand, direct  measurements of the expected top-heavy IMFs
in the heavily dust-enshrouded  compact starbursts where extreme CRDRs
would  reside are  impossible.  However  any cases  where  the ``end''
products, i.e.  the stellar populations, of high-$\rm \dot\rho _{sfr}$
SF  events  can be  observed  and their  IMF  determined  such as  the
Galactic Center (GC) and local  Ultra Compact Dwarfs (UCDs) allow such
direct  observational tests.  There  is indeed  strong evidence  of CR
heating  of  the   dense  molecular  gas  in  the   GC,  setting  $\rm
T_{kin}$$>$$\rm T_{dust}$,  and expected  strong effects on  the local
Jeans  mass and  ambipolar  diffusion timescales  (Yusef-Zadeh et  al.
2007).  Unfortunately, initial assertions about a top-heavy IMF in the
GC (e.g.   Nayakshin \& Sunyaev  2005) and the massive  Arches cluster
near it (Figer et al.  1999;  Stolte et al.  2002, 2005) have not been
corroborated  by later  studies (Kim  et  al.  2006;  Espinoza et  al.
2009; L\"ockmann et al.  2010). On the other hand some evidence exists
for a top-heavy IMF in the  nuclear starburst of M\,82 (e.g.  Rieke et
al.  1980;  1993),  and  a  significantly  enhanced  $\rm  U_{CR}$  is
certainly there ($\sim $500$\times$$\rm U_{CR,Gal}$, directly measured
via $\gamma$-ray  observations; Acciari et al.  2009)  to achieve this
via a CR-boosted $\rm M^{(*)}  _{ch}$ IMF mass scale.  Indeed for such
$\rm   U_{CR}$   values,   setting   $\rm   M^{(*)}_{ch}$$\sim   $$\rm
M_{J}(10^5\,cm^{-3})$                    yields                   $\rm
M^{(*)}_{ch}$(M\,82)$>$2\,M$_{\odot}$  (Fig.4).  However  the Galactic
Center  does offer  an example  of regions  with locally  boosted $\rm
U_{CR}$ values  that do not necessarily correspond  to top-heavy IMFs
in  their vicinity,  possibly because  such boosts  are too  short and
transient to systematically bias the local emergent IMF.

  Stronger evidence  for top-heavy IMFs exists in  UCDs, small stellar
systems with r$_e$$\sim  $(10--30)\,pc (half-light radius) where $\sim
$(10$^6$--10$^7$)\,M$_{\odot }$ of gas  have been converted into stars
at  $\rm \langle SFR\rangle  $$\sim $(10--100)\,M$_{\odot}$\,yr$^{-1}$
(see Dabringhausen et al.  2009 and references therein). Unlike the GC
where IMF studies are hindered by high extinction and tidal disruption
of clusters, and where any  high-$\rm \dot\rho _{sfr}$ SF episodes may
have  been short-lived  (and  thus could  not  significantly bias  its
average IMF towards  a top-heavy one), the IMFs in  UCDs are much more
amenable to systematic studies, while  their SF episodes have had $\rm
\langle \dot\rho _{sfr}(UCDs)\rangle
$$\ga  $$\rm \langle \dot\rho  _{sfr}(ULIRGs)\rangle$!  They  are thus
ideal targets for studying  the IMF resulting from completed high-$\rm
\dot\rho_{sfr}$ SF events.

The next  best thing besides studies  of the IMF itself  is studies of
the  ISM in  CRDRs from  which the  IMF emerges,  especially  those in
compact extreme starbursts  where $\rm \dot\rho _{sfr} $  is very high
(e.g.  ULIRGs) versus systems with  more distributed SF and much lower
$\rm \dot\rho _{sfr}$ values.   The unique observational signatures of
the very  warm dense  gas with high  ionization fractions  expected in
extreme  CRDRs   have  already  been  discussed  in   some  detail  by
Papadopoulos  (2010), and  the interested  reader is  refered  to this
work. In summary, in the era  of Herschel and the upcoming era of ALMA
the sole  main obstacle is  the degeneracy of extreme  CRDR diagnostic
with that of  X-ray Dominated regions (XDRs) induced in  the ISM by an
X-ray  luminous AGN.   Indeed, unless  an  X-ray luminous  AGN can  be
excluded by  other independent means (e.g.   deep X-ray observations),
it can be  difficult to tell extreme CRDRs and XRDs  apart in a simple
fashion  as even  exotic high  excitation species  such as  OH$^+$ and
H$_2$O$^+$, recently detected by  Herschel in the archetypal ULIRG/QSO
Mrk\,231 and attributed to XDRs (van  der Werf et al.  2010), could be
partly due  to CRDRs. In recent  work Meijerink et al.  2010 provide a
detailed discussion of diagnostics  lines in extreme CRDRs versus XDRs
and the key observations that can separate them.

 Finally it  is worth  noting that X-rays  and XDRs will  have similar
effects as CRs on the Jeans  mass and the IMF of UV-shielded dense gas
regions  as  both  can   volumetrically  warm  such  regions  to  high
temperatures.   This  was pointed  out  recently  by  Bradford et  al.
(2009) for the ISM of a distant QSO, and constitutes an omitted aspect
of AGN feedback on star formation.  This has been studied recently and
shown to lead to a top-heavy IMF (Hocuk \& Spaans 2010).  XDRs besides
elevating $\rm  T_{k}$(min) to much  higher values than in  the Galaxy
they also  yield $\rm T_{kin}$$\gg$$\rm  T_{dust}$ (and thus  they can
also ``erase''  the EOS inflection  point), though unlike  CRDRs their
influence on the  ISM is necessarily limited by  the $\sim $$\rm 1/R^2
_{AGN}$ geometric factor ($\rm R_{AGN}$ the distance of the irradiated
gas from the~AGN).
\vspace*{-0.5cm}

\section{Conclusions}

We have conducted new, detailed calculations on the thermal balance of
UV-shielded dense gas in Cosmic-Ray-Dominated-Regions (CRDRs), the ISM
phase  where the  initial  conditions  of star  formation  are set  in
galaxies, in order  to examine in detail the  effects of extreme CRDRs
on the  mass scale of  young stars and  the IMF recently  suggested by
Papadopoulos (2010).  Our results can be summarized as follows:

1. We confirm the high temperatures  of dense gas cores in the extreme
   CRDRs expected in ULIRGs and  all systems with large star formation
   rate densities ($\rm \dot\rho _{sfr}$), albeit at lower levels than
   the original study.

2. These  CR-induced higher gas  temperatures will  lead to  $\sim $10
   times larger  Jeans masses across the entire  density range typical
   in such regions, and thus higher charecteristic mass of young stars
   in the ISM of galaxies with high $\rm \dot\rho _{sfr}$.

3. The CRs,  by decoupling the thermal  state of the gas  from that of
   the  dust  in  CRDRs  keep  $\rm T_k$$\gg$$\rm  T_{dust}$  for  all
   densities typical of molecular clouds.  This effectively ``erases''
   the so called inflection point  of the Effective Equation of State,
   especially in extreme  CRDRs, inevitably altering the fragmentation
   properties of the  gas.  This effect, along with  the expected much
   higher gas  ionization fractions, necessitates  a search for  a new
   EOS, appropriate for CRDRs.

4. These new CR-induced ISM conditions  provide a new paradigm for all
  high-density star formation in the  Universe, with CRs as the key SF
  feedback mechanism,  operating unhindered by extinction  even in the
  most dust-enshrouded starbursts.

5. The  resulting $\rm  \dot\rho _{sfr}$-dependent  IMF, yields:  a) a
  natural   bimodal  behaviour   for   merger-driven/starburst  versus
  disk/regular-SF galaxies,  and b) integrated  galactic IMFs (IGIMFs)
  that  depend on their  star formation  history, both  with important
  consequences in galaxy formation and evolution.

6. IMF studies of stellar systems formed under extremely high $\langle
  \rm \dot\rho _{sfr}\rangle$ values such as Ultra Compact Dwarfs, and
  molecular line  observations of the  extreme CRDRs in  nearby ULIRGs
  can provide the full picture of CRDR properties and their effects on
  the  IMF.  Regarding the  latter, the  now spaceborn  Herschel Space
  Observatory, and the upcoming ALMA hold particular promise.

\section*{Acknowledgments}

We would like to thank Pavel Kroupa for numerous useful conversations,
especially  regarding the  IMF of  UCDs, and  Joerg  Dabringhausen for
pointing  out his  relevant work.   We  also thank  Andrew Strong  for
discussing  uncertainties regarding  CR transport  in  dense molecular
clouds, and the referee for comments regarding the CR energy densities
and their effect  on the IMF of M\,82.   {\bf Padelis P.  Papadopoulos
would like  to dedicate  this work  to the memory  of his  dear friend
Yannis Bakopoulos,  an excellent mathematician,  fellow scientist, and
committed social activist who died on 6th of December 2010.}

\newpage

\onecolumn
\appendix

\section{The cooling and heating functions for  UV-shielded cores}

The CR heating rate used in this work is

\begin{equation}
\rm \Gamma_{\mathrm{CR}}=1.95\times10^{-28}n_{\mathrm{H}}
 \left(\frac{\zeta_{CR}}{1.3\times10^{-17}\,s^{-1}}\right)\, ergs\, cm^{-3}\,s^{-1}
\end{equation}

\noindent
(Wolfire et  al. 1995) where  $\rm \zeta _{CR}$$\propto  $$\rm U_{CR}$
being the  Cosmic Ray  ionization rate (in  s$^{-1}$), with  a adopted
Galactic  value  of  $\rm  \zeta  _{CR}$=5$\times$10$^{-17}$  s$^{-1}$
(corresponding   to   $\rm   U_{CR}$=$\rm   U_{CR,Gal}$),   and   $\rm
n_{H}$=2n(H$_2$) for fully molecular  gas cores. For an optically thin
OI  line  the   cooling  is  $\rm  \Lambda_{\mathrm{OI63}}=\chi_O  n_H
C_{\mathrm{lu}}E_{\mathrm{ul}}$, which becomes

\begin{equation}
\rm \Lambda_{\mathrm{OI63}}= 3.14\times10^{-14}\chi_O
\rm n_{\mathrm{H}}C_{\mathrm{ul}}\left(\frac{g_{\mathrm{u}}}{g_{\mathrm{l}}}\right)
\rm \exp{\left(-227.72/T_{\mathrm{k}}\right)}\,ergs\,
\end{equation}

\noindent
where      $g_{\mathrm{u}}$=3     and      $g_{\mathrm{l}}$=5,     and
$\chi_{\mathrm{O}}$=[O/H]  being the  abundance of  oxygen  not locked
onto    CO    ($\chi_{\mathrm{O}}$$\sim$4.89$\times$10$^{-4}$).    The
collisional de-excitation coefficient is given by 

\begin{equation}
\rm C_{\mathrm{ul}}=n_{\mathrm{H}} 10^{0.32\log{T_{\mathrm{k}}}-10.52}=
\rm     3.02\times10^{-11}n_{\mathrm{H}}T_{\mathrm{k}}^{0.32}\,cm^{-3}\,s^{-1},
\end{equation}

\noindent
(Liseau et al. 1999). Thus finally Equation A2 becomes,

\begin{equation}
\rm \Lambda_{\mathrm{OI63}}=2.78\times10^{-28}n_{\mathrm{H}}^2T_{\mathrm{g}}^{0.32}
\rm \exp{\left(-227.72/T_{\mathrm{k}}\right)}\,ergs\,cm^{-3}\,s^{-1}.
\end{equation}

\noindent
  At lower densities  ($n_{\mathrm{H}}<$10$^{5}$ cm$^{-3}$) and strong
CR fluxes,  a small  fraction of  carbon remains in  the form  of CII,
acting as a coolant with

\begin{equation}
\rm \Lambda_{\mathrm{CII}}=1.975\times10^{-23}n_{\mathrm{H}}^2\chi_{\mathrm{CII}}
\rm \exp{\left(-92.2/T_{\mathrm{k}}\right)}\,ergs\,cm^{-3}\,s^{-1},
\end{equation}

\noindent
computed in a similar fashion as the OI(63\,$\mu $m) line cooling, and
for a  fully transparent medium.  Gas-grain accommodation  can heat or
cool   the   gas   depending   on   the   difference   between   their
temperatures (Burke \&  Hollenbach 1983). It can be expressed as

\begin{equation}
\rm \Gamma_{\mathrm{acc}}=4.0\times10^{-12}\alpha n_{\mathrm{H}}
\rm n_{\mathrm{d}}\sigma_{\mathrm{g}}\sqrt{T_{\mathrm{k}}}
\rm \left(T_{\mathrm{k}}-T_{\mathrm{dust}}\right)
\end{equation}

\noindent
 where  $n_{\mathrm{g}}$ is  the  number density  of  dust grains  and
$\sigma_{\mathrm{g}}$ is  the grain cross-section  (cm$^{-2}$). If the
gas-to-dust    ratio   is    100,   the    dust   mass    density   is
$\rho_{\mathrm{d}}$=3.5\,g\,cm$^{-3}$,   and    a   dust   radius   of
$a=$~0.1~$\mu$m then

\begin{equation}
\rm n_{\mathrm{d}}=2.2\times 10^{-2}\times\mu_{\mathrm{H}}
\rm n_{\mathrm{H}}/(4/3\pi\rho_{\mathrm{d}}a^{3})=7.88\times10^{-12}n_{\mathrm{H}}.
\end{equation}

\noindent
 For the accommodation factor $\alpha $ we follow the treatment by Groenewegen (1994)
 where

\begin{equation}
\rm \alpha=0.35\exp{\left(-\sqrt{\frac{T_{\mathrm{dust}}+T_{\mathrm{k}}}{500}}\right)},
\end{equation}

\noindent
which we  set to  the maximum value  $\alpha$=0.35 (i.e.   maximum gas
cooling from gas-dust interaction).  Thus the mainly cooling term (for
the ISM conditions  explored here) due to the  gas-dust interaction in
Equation A6 becomes

 \begin{equation}
\rm \Lambda_{\mathrm{gd}}=3.47\times10^{-33}n_{\mathrm{H}}^{2}
\rm \sqrt{T_{\mathrm{k}}}\left(T_{\mathrm{k}}-T_{\mathrm{dust}}\right) ergs\,cm^{-3}\,s^{-1}.
\end{equation}

\noindent
Finally based on our UCL-PDR code outputs we parametrize the cooling 
due to the CO rotational transitions as

\begin{equation}
\rm \Lambda_{\mathrm{CO}}=4.4\times10^{-24}\left(\frac{n_{\mathrm{H}}}{10^4\,cm^{-3}}\right)^{3/2}
\rm \left(\frac{T_{\mathrm{k}}}{10\,K}\right)^2
\rm \left(\frac{\chi_{\mathrm{CO}}}{\chi_{\mathrm{[C]}}}\right)\,ergs\,cm^{-3}\,s^{-1},
\end{equation}

\noindent
where                              we                              set
$\chi_{\mathrm{CO}}/\chi_{\mathrm{[C]}}$=(0.97,0.98,0.99,1.0)       for
densities of 5$\times$10$^{3}$, 10$^{4}$, 10$^{5}$, 10$^{6}$ cm$^{-3}$
respectively (with the value of  =1 corresponding to all carbon locked
onto CO).

Solutions of  Equation 2 in the  main text for  densities spanning the
range  present in  star-forming  H$_2$ clouds  in  galaxies, and  $\rm
U_{CR}$=(0.2,   1.0,   10,   10$^2$,  10$^3$,   10$^4$)$\times   $$\rm
U_{CR,Gal}$ are then obtained numerically and are shown in Figure 1.

\label{lastpage}

\end{document}